\documentclass[twocolumn,showpacs,preprintnumbers,amsmath,amssymb,prb,superscriptaddress]{revtex4}
\usepackage{graphicx}
\usepackage{dcolumn}
\usepackage{bm}
\usepackage{epstopdf}

\begin{document}
\title{Estimates of Effective Hubbard Model Parameters for C$_{20}$ isomers}
\author{Fei Lin}
 \affiliation{Department of Physics, University of Illinois at Urbana-Champaign, Urbana, Illinois 61801, USA}
\author{Erik S. S{\o}rensen}
 \affiliation{Department of Physics and Astronomy, McMaster University, Hamilton, Ontario,
 Canada L8S 4M1}
 \date{\today}

\begin{abstract}
We report on an effective Hubbard Hamiltonian approach for the study of electronic correlations in C$_{20}$ 
isomers, cage, bowl and ring, with quantum Monte Carlo and exact diagonalization methods. The tight-binding hopping
parameter, $t$, in the effective Hamiltonian is determined by a fit to density functional theory calculations, 
and the on-site Coulomb interaction, $U/t$,  is determined by calculating the isomers' affinity energies, 
which are compared to experimental values. For the C$_{20}$ fullerene cage we estimate $t_{\rm cage}\simeq 0.68-1.36$ eV
and $(U/t)_{\rm cage} \simeq 7.1-12.2$. The resulting effective Hamiltonian is then used 
to study the shift of spectral peaks in the density of states of neutral and one-electron-doped 
C$_{20}$ isomers. Energy gaps are also extracted for possible future comparison with 
experiments. 

\end{abstract}


\maketitle

\section{Introduction}
\begin{figure}[t]
\begin{center}
\begin{tabular}{c}
\includegraphics[clip,width=3.5cm]{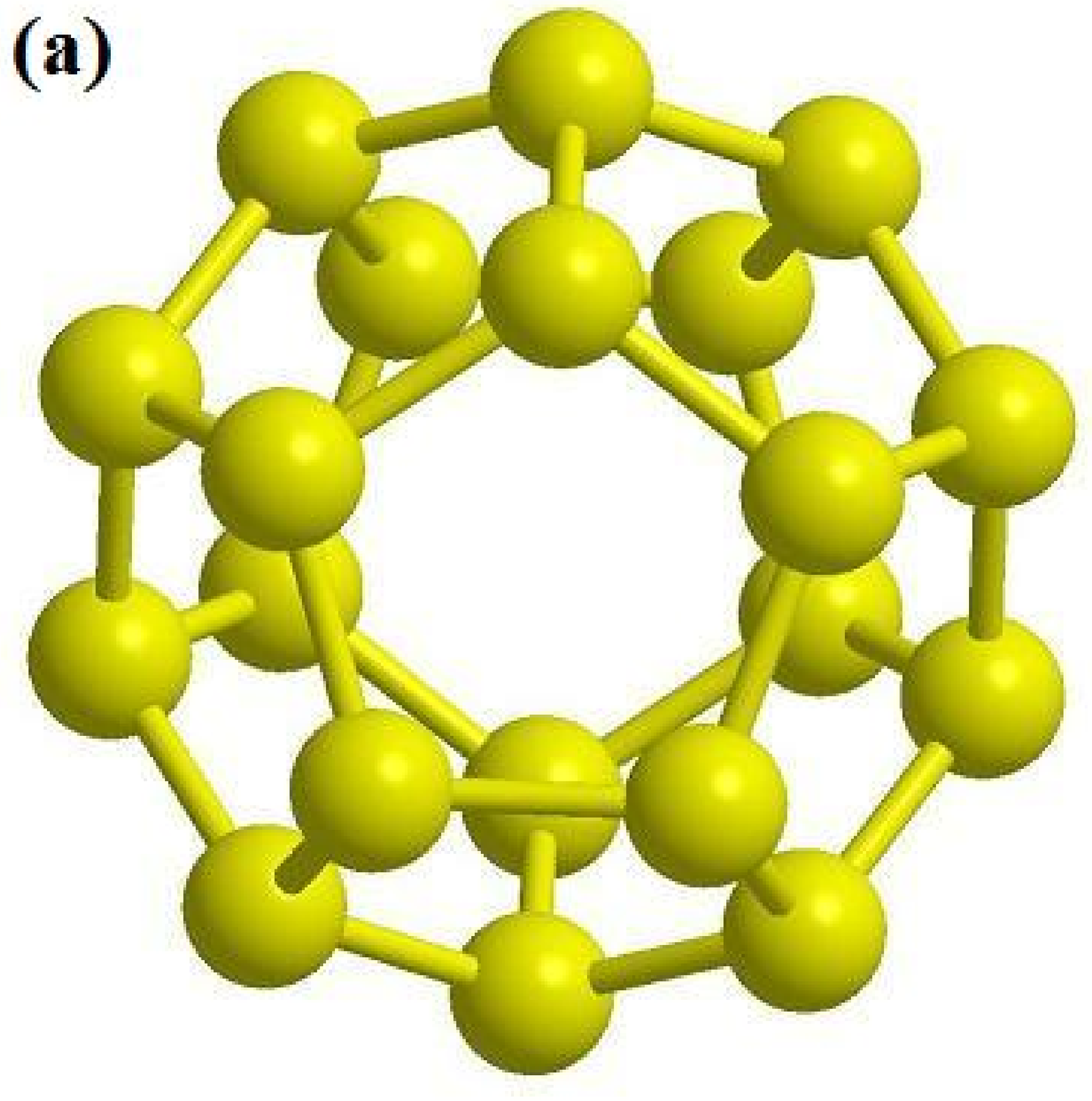} \\
\includegraphics[clip,width=3.5cm]{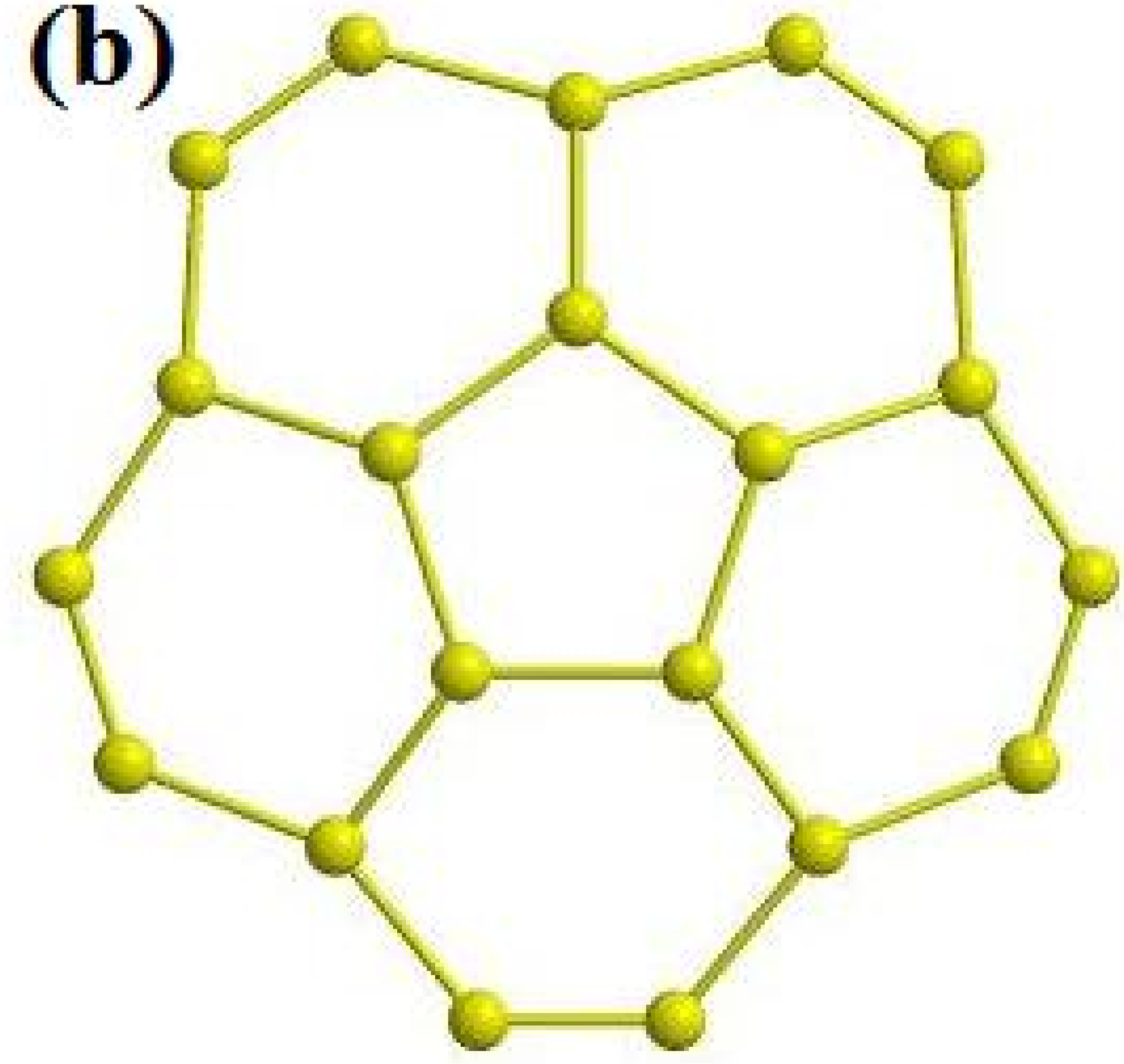} \\
\includegraphics[clip,width=3.5cm]{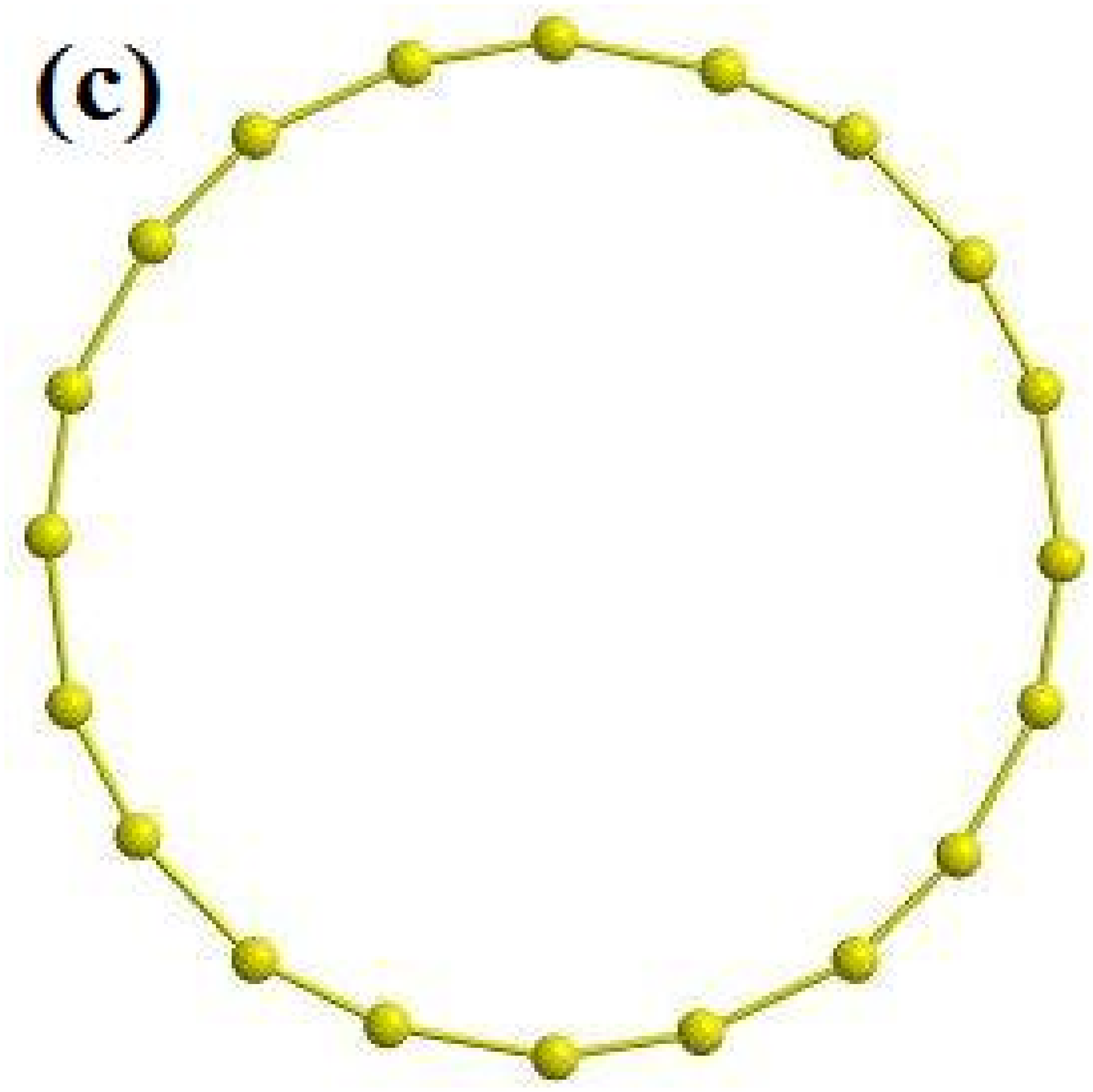}  \\
\end{tabular}
\caption{(Color online.) Molecular structures of C$_{20}$ isomers: (a) cage, (b) bowl, and (c) ring.}
\label{isomerpic}
\end{center}
\end{figure}

The successful synthesis of gas phase C$_{20}$ molecules displaying the dodecahedral fullerene
cage structure \cite{prinzbach00} has induced considerable research interest in this smallest 
fullerene molecule, partly because of the superconducting property of electron-doped 
C$_{60}$ materials \cite{hebard91} in the same fullerene family, and partly because 
previous theoretical speculations \cite{grossman95, murphy98, raghavachari93} about the 
existence of such a cage molecule were confirmed. In the experiment,\cite{prinzbach00} 
three major C$_{20}$ isomers, of cage, bowl and ring structures, were produced. See Fig. \ref{isomerpic} 
for the molecular structures. Photoelectron spectra (PES) were also measured for all three brominated and 
one-electron doped isomers (C$_{20}^{-}$). Affinity energies ($AE$, see definition below) were 
then extracted from the PES figure, giving $AE_{\textrm{cage}}=2.25$ eV, $AE_{\textrm{bowl}}=2.17$ eV, 
and $AE_{\textrm{ring}}=2.44$ eV, respectively. Since PES spectra reflect both the isomers 
geometric character and the strength of electronic correlation inside the molecules, 
it is a unique opportunity to investigate the interplay between geometry and electronic 
correlations in the three isomers. Here, we shall do this by estimating the parameters
in an effective Hubbard model description of these isomers, parameterized in terms of
the on-site repulsion, $U$, and hopping integral, $t$. We find that for the 
geometry with the highest curvature, the fullerene cage, correlations effects as measured
by $U/t$ are relatively important.
Experiments have also shown evidence for
solid phases of C$_{20}$ fullerene cage~\cite{wang01, iqbal03} further emphasizing the need for
understanding strong correlation effects in this isomer.

The possibility of superconductivity arising from a purely electronic mechanism in the C$_{60}$ fullerenes
was suggested in the early 90's~\cite{kivelson91a} and supported by perturbative calculations~\cite{kivelson91a,kivelson91b} starting
from a one-band Hubbard model.
In this picture superconductivity arises from strong correlation effects since the many-body energy levels favor
electrons residing on the same molecule as opposed to different molecules, resulting in a negative pair-binding energy.
However, extensive numerical work~\cite{lin05a} has shown that the pair-binding energy in C$_{60}$ materials
is likely {\it positive} for $U/t\leq 4.5$ with
larger values of $U/t$ inaccessible due to an increasingly severe sign-problem.
Comparatively, C$_{20}$ has a larger curvature and correlation effects measured in terms of $U/t$ are therefore likely much more important since
the curvature will decrease $t$ and thereby increase $U/t$.
We have previously studied electron correlation effects in the C$_{20}$ fullerene cage \cite{lin07} starting from a 
one-band Hubbard model defined in Eq. (\ref{exthubmd}). For the whole range of $0<U/t<\infty$, we find that the 
pair-binding energy is always {\it positive} ruling out the possibility of superconductivity induced solely by an electronic 
mechanism. With increasing $U/t$ there is, however, a very interesting metal-insulator 
transition predicted~\cite{lin07} to occur around
$U/t\sim 4.1$ in molecular solids formed with the $C_{20}$ fullerene.
Experiments have shown evidence for such
solid phases of C$_{20}$.~\cite{wang01, iqbal03}
We expect that most of the strong correlations effects are fully captured by a one-band Hubbard model
with a single {\it uniform} hopping integral. In order to develop predictive models of C$_{20}$ materials and to determine
whether such materials will display metallic or insulating behavior
it is then crucial to estimate how large $U/t$ is for the C$_{20}$ molecule, which 
is the main purpose of this paper.

Mean field density functional theory (DFT) calculations have yielded conflicting predictions of 
the relative stability of C$_{20}$ isomers. This suggests that electron correlations, which are 
only approximately treated in the DFT calculations, could be very important in the prediction of the electronic
structure of C$_{20}$ isomers even though the geometry of the different isomers is fully
captured in the DFT calculations.
We are, 
therefore, inspired to study the isomers with an effective Hamiltonian approach, where 
a Huckel Hamiltonian is complemented with an on-site electron correlation term, i.e., 
the one-band Hubbard model defined on a single C$_{20}$ molecule as
\begin{equation}
H=-\sum_{\langle
ij\rangle\sigma}t_{ij}(c^{\dagger}_{i\sigma}c_{j\sigma}+h.c.) +U\sum_i
n_{i\uparrow}n_{i\downarrow}+\varepsilon_p\sum_i n_i,
\label{exthubmd}
\end{equation}
where $c^{\dagger}_{i\sigma}$ ($c_{i\sigma}$) is an electron
creation (annihilation) operator with spin $\sigma$ on site $i$, indices $i,j$ run
over 20 sites of the isomers, $t_{ij}$ is the hopping integral between nearest neighbor 
(NN) carbon atoms $i$ and $j$, $U$ is the on-site electron correlation,
$n_i=n_{i\uparrow}+n_{i\downarrow}$ is the number of electrons on
site $i$, and $\varepsilon_p$ is the on-site electronic energy due to the core ion and electrons 
in the carbon atom. According to Ref. \onlinecite{harrison89}, we set $\varepsilon_p=-8.97$ eV. 
This on-site electronic energy term is a constant as long as the total electron number in the 
molecules is fixed, and is not important in calculating, e.g., the pair-binding energy, 
\cite{lin07} but needs to be taken into account when calculating the affinity energy of 
the molecules, defined as
\begin{equation}
AE=E(20)-E(21),
\end{equation} 
where $E(N)$ is the internal energy of the molecule filled with $N$ electrons from $2p$ atomic orbitals.

The idea of fitting a tight-binding Huckel Hamiltonian to DFT energy levels for a fullerene 
molecule is not new and has been employed in, e.g., Ref. \onlinecite{satpathy86, 
manousakis91}. However, inclusion of an on-site Coulomb interaction in the tight-binding 
model for the fullerene molecule has not been very well studied due to the difficulty of performing
reliable calculations in the presence of such a term. In particular, the question 
of what value the on-site interaction $U$ should take has not been answered for C$_{20}$ isomers.
In light of the metal-insulator transition predicted~\cite{lin07} to occur around $U/t=4.1$ a correct
determination of $U/t$ is crucial for modeling C$_{20}$ based materials.
Addressing this question is 
our main goal here. The paper is organized as follows.

First, since we want the tight-binding $t$ term (Huckel Hamiltonian) to reflect the geometric 
character of each of the isomers, which is contained in the DFT energy levels, we fit 
the Huckel energy levels to energy levels obtained from DFT. This allows us to determine the effective value
of $t$ for the three isomers. Here we assume {\it uniform} hopping integrals $t$ inside the 
cage and bowl molecules for the 
Huckel Hamiltonian, 
although within the DFT approach for the bowl isomer the
hopping integrals are {\it not} uniform \cite{raghavachari93} even 
in the absence of any Jahn-Teller distortion. 
We expect this simplification for the Huckel approach
to be of only minor importance for the bowl and the cage.
However, for the ring isomer, DFT calculations show that the ring structure is 
dimerized \cite{raghavachari93} with 
alternating long and short bonds between NN carbon atoms, which leads to a filled highest occupied molecular orbital (HOMO), i.e. an insulating molecule, in contrast to 
the uniform bonding case, where the HOMO is not completely filled (the molecule is metallic). We, therefore, consider two different 
hopping integrals for the ring Huckel Hamiltonian. As mentioned, inhomogeneous hopping integrals are always  
considered in the DFT calculations. We then study the effect of electron correlation in the 
neutral isomers by calculating the single-particle excitation spectra with different 
correlation strengths ($U/t=2,3,4,5$) and show the difference between quantum Monte Carlo (QMC) and DFT spectra 
for the neutral isomers. To complete our effective Hamiltonian approach, we proceed to 
estimate the on-site Coulomb interaction strength $U$ by calculating the electron affinity 
energies of the isomers and comparing them to the experimental values. This allows us to determine 
$U/t$  for each isomer and  we can then study the effect of one-electron doping on the single-particle 
excitation spectra with QMC simulations. We compare the resulting QMC spectra to results obtained from
DFT. These spectra should be directly comparable to possible future experimental PES and inverse PES spectra. 
\begin{table}[t]
  \centering
  \begin{tabular}{|c|ccc|ccc|}
    \hline\hline
    &  & DFT &  &  & Huckel ($U=0$)& \\
    \hline
  & $E$ (eV)  & $D$  & $N_e$  & $E/t$ & $D$  & $N_e$ \\
    \hline
 cage &  1.72766  & 3    &   0    &   2.23607  & 3  & 0 \\
      &  0.37579   & 1    &   0    &   2.0      & 4  & 0 \\
      & -0.98560   & 4    &   2    &   0.0      & 4  & 2 \\ 
      & -6.65895   & 5    &   10   &  -1.0      & 5  & 10 \\
    \hline
 bowl & -2.15325   & 1    &   0    &   1.0      & 1  & 0 \\
      & -2.50728   & 1    &   0    &   0.47725  & 2  & 0 \\
      & -5.00612   & 1    &   2    &  -0.73764  & 2  & 4 \\
      & -5.02272   & 1    &   2    &  -0.77748  & 2  & 4 \\
     \hline
 ring & -6.96644   & 1    &   0    &   0.63303  & 2  & 0 \\
      & -7.54251   & 1    &   0    &   0.14400  & 1  & 0 \\
      & -8.45383   & 1    &   2    &  -0.14400  & 1  & 2 \\
      & -8.94745   & 1    &   2    &  -0.63303  & 2  & 4 \\
  \hline\hline
  \end{tabular}
\caption{Comparison of energy levels around the Fermi energy from DFT calculation and 
Huckel Hamiltonian on the C$_{20}$ isomers. The energy from DFT is in units of eV.
$D$ is degeneracy degree of the corresponding energy level, and $N_e$ is the number 
of electrons occupying these energy levels. For each isomer, the energy levels are 
listed in descending order.} \label{dfthuckel}
\end{table}

\section{Estimation of the Huckel hopping integral, $t$}~\label{sec:huck}
Different molecular geometries of C$_{20}$ isomers determine different NN hopping 
integrals $t$ in the Huckel description of molecules. In this section, we estimate 
approximate values of $t$'s for the cage, bowl and ring structures of C$_{20}$ molecules,
by comparing tight-binding Huckel energy diagrams ($U=0$) with the energy levels from DFT 
calculations. In DFT calculations for the cage and bowl,
$2s$, $2p_x$, $2p_y$, and $2p_z$ atomic orbitals are considered, 
and the calculations are performed with the widely-used ABINIT code. \cite{abinitpg} 
Since there are 4 orbitals per carbon atom in the DFT calculation, the resulting energy 
levels are a mixture of $\sigma$ and $\pi$ bonds. The $\sigma$ bonds have either very low
or very high energies. Energy levels around the Fermi energy mainly consist of $\pi$ bonds, which 
we use to construct the Huckel Hamiltonian. The comparison of energy levels between DFT 
and tight-binding Huckel Hamiltonian should, therefore, be made around Fermi energy. Table 
\ref{dfthuckel} shows such a comparison. To fit the Huckel hopping parameters $t$, we set 
equal the {\it smallest} energy gap of the tight-binding Huckel Hamiltonian with an equivalent gap in
the DFT results. The degeneracy of the levels are here important and since the variation in the hopping integrals
in the DFT approach in some cases will split levels the equivalent gap in the DFT approach is {\it not necessarily}
the smallest gap. In studies of strong correlation effects
it is highly desirable to use the simplest possible model. Since for the fullerene cage the DFT and
the tight-binding model with {\it uniform} hopping at $U=0$ both predict a half filled HOMO level and  metallic behavior, we have not found it necessary
to include variations in the hopping integral. The same is true for the bowl where a filled HOMO level is found.
However, as previously mentioned, for the ring the two approaches disagree and we are therefore forced to
include a variation in the hopping integrals for the ring as we describe in detail below.

We begin by discussing the fullerene cage.
Here we find a half-filled 4-fold degenerate HOMO level and hence a metal in both the DFT ($-0.98560$ eV) and tight-binding ($0t_{\rm cage}$) results. 
In the tight-binding approach the first excited level is 4-fold degenerate at $2t_{\rm cage}$. 
Assuming that this 4-fold level
is split into a high lying 3-fold and a lower lying non-degenerate level due to slight variations in the effective hopping
integrals in the DFT approach, the equivalent gap in the DFT approach should be from the 4-fold level at $-0.98560$ eV to some
average of the 3-fold and non-degenerate excited levels. In the extremal case we neglect the non-degenerate level and
we then see that this gap should be close to the $(1.72766+0.98560)$ eV$=2.71326$ eV from which  we infer that
$t_{\rm cage}\simeq 1.36$ eV. Here we have taken the gap in the DFT calculations to go from the 4-fold degenerate level
to the 3-fold degenerate level. Given the splitting of the excited levels in the DFT approach it seems plausible that this
is a maximal value for the gap. However, in the absence of analytical results for the splitting of the levels in the DFT approach we note
that this is not an exact bound. The estimate $t_{\rm cage}\simeq 1.36$ eV is then likely an upper bound on the hopping integral.
An extreme {\it lower} bound on $t_{\rm cage}$ can be obtained if we use the smallest gap of $(0.37579+0.98560)$ eV$=1.36139$ eV in the DFT approach,
yielding $t_{\rm cage}^{\rm min}\simeq 0.68$ eV. Given the large splitting of the non-degenerate level from the 3-fold level this
lower bound seems very unlikely to be attained and a more realistic value for $t_{\rm cage}$ is likely close to the upper bound
of $1.36$ eV which we mainly focus on in the following.

We next turn to the bowl where one in both the DFT and tight binding approach ($U=0$) finds a filled HOMO level
and hence an insulator.
For the bowl we compare the tight-binding gap of $(0.47725+0.73764)t_{\rm bowl}=1.21489t_{\rm bowl}$ from
a 2-fold degenerate level to another
2-fold degenerate level with the 
$(-2.15325+5.02272)$ eV=$2.86947$ eV gap in the DFT results yielding $t_{\rm bowl}\simeq 2.36$ eV.
As for the fullerene cage, we have here assumed that both 2-fold degenerate levels are slightly split in the DFT approach and we 
have used the largest reasonable value for the equivalent gap in the DFT approach.
Following the discussion of the hopping integral for the fullerene cage we again expect the estimate
$t_{\rm bowl}\simeq 2.36$ eV to be an upper bound. In this case, a reasonable {\it lower} bound on the hopping integral
can be obtained by taking the smallest gap in the DFT approach of $(-2.50728+5.00612)$ eV=$2.49884$ eV, resulting in $t_{\rm bowl}^{\rm min}\simeq 2.06$ eV, a relatively
modest variation from our previous maximal estimate.

Finally we turn to a discussion of the ring molecule. As mentioned, in this case a {\it uniform} tight binding model would predict a half-filled HOMO level
where as the DFT approach shows a filled HOMO level. We are therefore forced to include a staggering in the hopping integrals in the tight-binding approach.
For the ring DFT calculation, we first generate a pseudo potential \cite{atompaw} for the carbon atom, with 4 electrons ($1s^22s^2$) in the core state and 2 
electrons ($2p^2$) in the valence state. The resulting pseudo potential are then fed to the ABINIT \cite{abinitpg} to carry out DFT calculations on the 2 valence 
orbitals. On the tight-binding calculation side, we use 2 hopping integrals $t_l$ and $t_s$ to represent the hopping integrals of the alternating long 
($b_l=2.609$ Bohr)~\cite{raghavachari93} and short ($b_s=2.260$ Bohr)~\cite{raghavachari93} bonds, respectively. Let the average of the two bonds be $b=(b_l+b_s)/2=2.435$ Bohr. 
We then parametrize \cite{lin07} the two hopping 
integrals as $t_x/t_{\textrm{ring}}=1- (b_x-b)/b$, where $x=l, s$, and $t_{\textrm{ring}}$ is the average hopping integral.
With this 
parameterization we find $t_l=0.928t_{\textrm{ring}}$ and $t_s=1.072t_{\textrm{ring}}$. From Table \ref{dfthuckel} we see that both DFT and tight-binding 
calculations now give an insulating molecule. If we again compare the tight-binding gap of $(0.14400+0.14400)t_{\textrm{ring}}=0.288t_{\textrm{ring}}$ from 
a non-degenerate level to another non-degenerate level with the gap between two non-degenerate levels in the DFT results of
$(-7.54251+8.45383)$ eV=$0.91132$ eV, resulting in an average $t_{\rm ring}\simeq 3.16$ eV.

We see that, compared to the other isomers, the hopping integral is {\it significantly smaller} in the fullerene cage isomer due to the large 
curvature of the molecule that reduces the NN overlap of the $2p$ orbitals. 
Consequently, 
$t_{\textrm{cage}}$ for the C$_{20}$ fullerene is also significantly smaller than what was found for the much
bigger and less curved C$_{60}$ where one observes
$t_{C_{60}}=2.50$ eV according to our calculation with ABINIT as well as Ref. \onlinecite{manousakis91} 
or $2.72$ eV according to Satpathy's early calculation.\cite{satpathy86} 
It is also noteworthy that our estimate is likely an upper bound on $t_{\rm cage}$.
We also note that $t_{\rm ring}\sim 2t_{\rm cage}$, which reflects the fact that $t_{\rm ring}$ is 
the hopping integral of four $2p$ orbitals as apposed to $t_{\rm cage}$ being the hopping integral of two $2p_{\pi}$ orbitals.

\section{DOS from DFT calculation}~\label{sec:dos}
\begin{figure}[t]
\begin{center}
\includegraphics[clip,width=8cm]{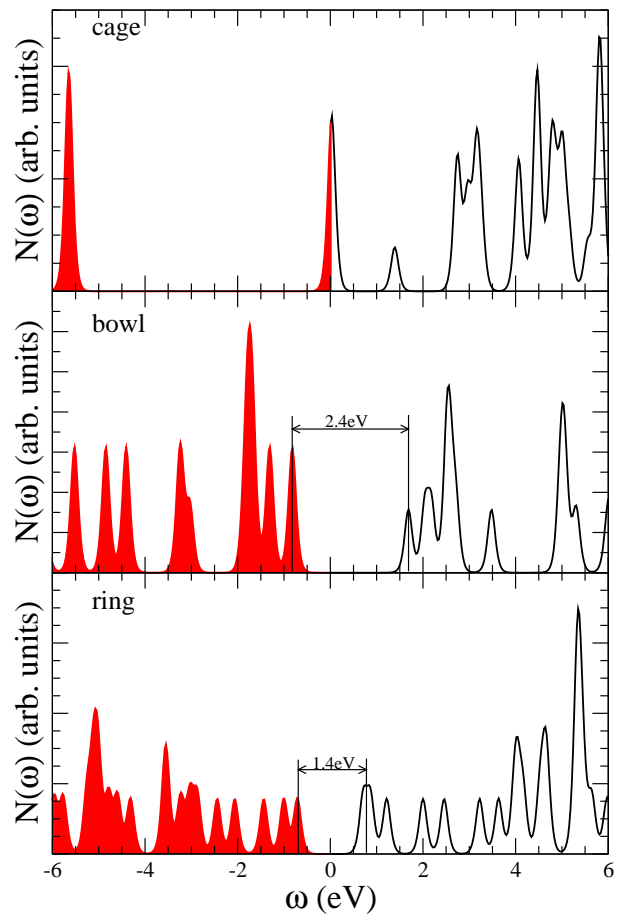}
  \caption{(Color online.) DOS from DFT for neutral cage, bowl and ring C$_{20}$ isomers. 
  Fermi energy is located at $\omega=0$. Shaded areas are occupied by electrons.}
  \label{dftdos}
\end{center}
\end{figure}
Before we study the effect of electronic correlation with QMC, we calculate the density of states 
(DOS) for the neutral C$_{20}$ isomers within DFT, shown in Fig. \ref{dftdos}. We see that both the
bowl and ring are 
insulators, with energy gaps $\Delta_{\textrm{bowl}}^{\textrm{DFT}}=2.4$ eV and 
$\Delta_{\textrm{ring}}^{\textrm{DFT}}=1.4$ eV for the neutral molecule. The fullerene cage isomer is, however, metallic. 
Electron correlations are typically underestimated in DFT calculations and it therefore
seems reasonable to assume that the inclusion of on-site 
electronic correlations would enlarge the gap, eventually turning the cage C$_{20}$ isomer into an 
insulator. Such an effect was shown to occur in Ref.~\onlinecite{lin07}.

\section{Effect of on-site correlation}\label{sec:effect}
\begin{figure}[t]
\begin{center}
\includegraphics[clip,width=8cm]{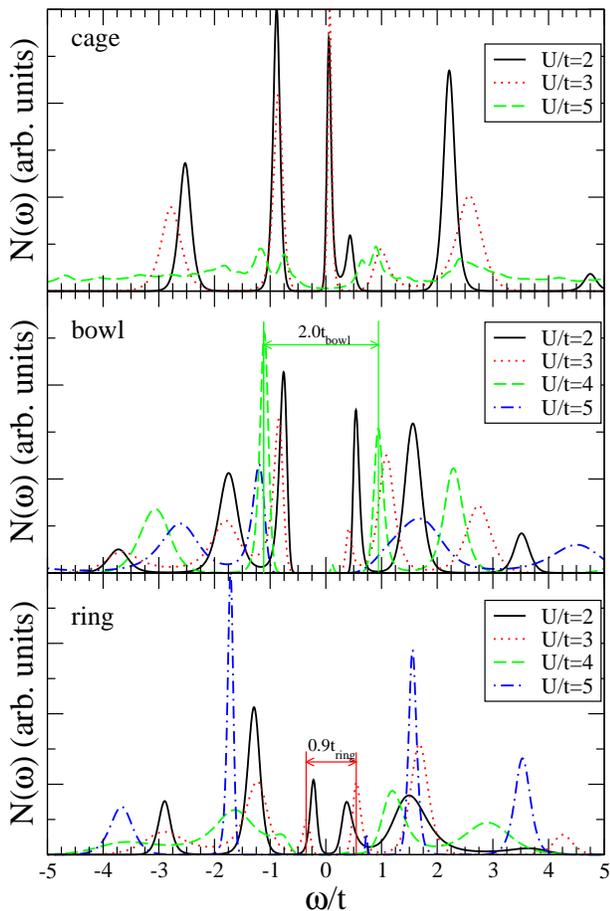}
  \caption{(Color online.) Evolution of the DOS with different $U/t$ values for neutral C$_{20}$ isomers: cage, bowl and ring. $U/t=5$ for 
    cage isomer is from exact diagonalization. \cite{lin07} Others are from QMC simulations. Fermi energies are located at $\omega=0$.}
  \label{qmcdos}
\end{center}
\end{figure}

As mentioned, DFT calculations typically underestimate electron correlations. However, with
the overlap integral, $t$, determined we can attempt to more closely describe the physics in the vicinity  of the Fermi energy by
using an effective one-band Hubbard model, Eq. (\ref{exthubmd}),  to account for
the electron correlations.
To perform such a study of on-site electronic correlation on the C$_{20}$ isomers we use
the standard QMC algorithm. \cite{hirsch85, white89, jarrell96} In regimes where the sign problem in this approach
renders results unobtainable we have supplemented these results by exact diagonalization (ED) results.
This allows us to determine the DOS
as a functions of $U/t$. The DOS is calculated for 
each of the neutral isomers for $U/t=2, 3, 4, 5$. Our results are shown in Fig. \ref{qmcdos}. As expected, the energy gap increases with 
increasing $U/t$ values for the bowl and the ring. The dependence on $U/t$ is clearly non-trivial. For the fullerene
cage the initially metallic molecule becomes insulating with increasing $(U/t)_{\rm cage}$. For molecular
solids formed out of this isomer a metal-insulator transition is therefore expected around
$(U/t)_{\rm cage}=4.1$.~\cite{lin07}

In the next section we estimate the on-site Coulomb interaction $U$ 
using the affinity energy of the isomers. However, assuming that the on-site electronic
energy scale is $U\sim 10$ eV for all the isomers, \cite{gunnarsson92, kivelson92, mila04, herbut06} we can do a rough estimate of 
$U/t$ and hence estimate the gap for the neutral molecule from the above QMC results for
the DOS. 
Using this value for $U$ we find
that $U/t\sim 4.2,3.2$ for the bowl and ring isomers respectively 
using
$t_{\textrm{ring}}\simeq 3.16$ eV and the upper bound  $t_{\textrm{bowl}}\simeq 2.36$ eV. For 
these values of $U/t$ and from Fig. \ref{qmcdos}, we estimate energy gaps for 
the neutral bowl and ring isomers to be about $\Delta_{\textrm{bowl}}=2.0t_{\textrm{bowl}}=4.72$ eV 
and $\Delta_{\textrm{ring}}=0.9t_{\textrm{ring}}=2.73$ eV, respectively. 
For the cage isomer $(U/t)_{\textrm{cage}}$ should be around 7 using the upper bound  $t_{\textrm{cage}}=1.36$ eV. 
Unfortunately, the sign problem prevents us from calculating the DOS and hence the gap by QMC simulations for 
$(U/t)_{\textrm{cage}}>3$ and we have to resort to the much more time consuming ED approach.
Previously, using ED techniques, the DOS and gap at $(U/t)_{\textrm{cage}}=5$ has been determined~ \cite{lin07}
yielding a gap of 1.89 eV. (See Fig.~\ref{qmcdos}) We have repeated this calculation at $(U/t)_{\textrm{cage}}=10$
finding a gap of 7.67 eV. Linearly interpolating between these values we estimate the
gap for the fullerene cage at
$(U/t)_{\textrm{cage}}\sim 7$ to be $\Delta_{\textrm{cage}}=3.1t_{\textrm{cage}}=4.2$ eV.

\section{Estimate of the on-site interaction, $U/t$}~\label{sec:U}
\begin{figure}[t]
\begin{center}
\includegraphics[clip,width=8cm]{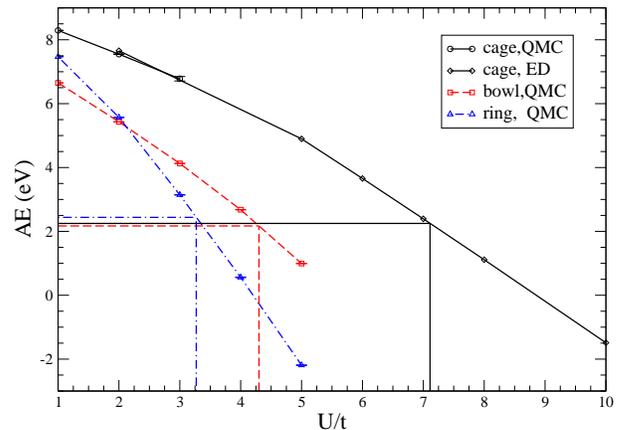}
  \caption{(Color online.) Variation of affinity energy with $U/t$ for a C$_{20}$ cage, bowl and ring, from
  QMC and ED calculations. The experimental $AE$ values are shown by the horizontal lines intercepting the 
  $AE \sim U/t$ curves. The corresponding $U/t$ values are shown by the vertical lines intercepting the 
  curves. Results are shown using the upper bound for $t$ for the cage and bowl.}
  \label{afnty}
\end{center}
\end{figure}

Electron correlations are important for an accurate calculation of the affinity energy, as shown by early QMC simulations. See, 
e.g., Ref. \onlinecite{barnett86}. Therefore, starting from the affinity energy, one can use QMC simulations to determine electron 
correlations, which in the case of Hubbard model are represented by the on-site interaction $U$. 
Since the electron affinity energies, $AE$, were measured experimentally,
\cite{prinzbach00} we can use these energies for a more refined estimate of the value of the
on-site effective Coulomb interaction $U$ for each of the isomers. Ground state energies of the isomers are calculated with
projection QMC technique \cite{white89} in the Hilbert space of fixed particle number $N$ and magnetization and converted to electron volts
using the previously determined estimates for $t$.
The affinity energy $AE=E(20)-E(21)$ is then subtracted and shown in Fig.
\ref{afnty} as a function of $U/t$. Results are shown using the upper bound for $t$ for the cage and bowl.
The sign problem prevents us from simulating larger $U/t$
values with QMC. Instead we use ED to obtain a series of $AE$ for $U/t=2, 5, 6, 7, 8, 10$
for the cage isomer. Note that, for the fullerene cage the electron affinity changes slope around
the expected critical value of $(U/t)_{\rm cage}=4.1$.~\cite{lin07}
We see that all 3 curves are approximately linear around the experimental 
affinity energy values \cite{prinzbach00} ($AE_{\textrm{cage}}=2.25$ eV, $AE_{\textrm{bowl}}=2.17$ eV,
and $AE_{\textrm{ring}}=2.44$ eV). Thus, a linear interpolation gives $U_{\textrm{cage}}=7.1t_{\rm cage}$, 
$U_{\textrm{bowl}}=4.30t_{\textrm{bowl}}$, and $U_{\textrm{ring}}=3.27t_{\textrm{ring}}$. 
If we for the fullerene cage perform a similar analysis using $t_{\rm cage}^{\rm min}=0.68$ eV we find
instead an even larger value for $(U/t)_{\rm cage}\simeq 12.2$. Likewise, we find for the bowl
using $t_{\rm bowl}^{\rm min}=2.06$ eV a value of $(U/t)_{\rm bowl}=4.9$.
                                                                                                               
The above results (Fig.~\ref{afnty}) show that for cage, bowl and ring isomers the on-site Coulomb interaction energies are 
$U_{\textrm{cage}}=7.1t_{\textrm{cage}}=9.67$ eV, 
$U_{\textrm{bowl}}=4.30t_{\textrm{bowl}}=10.15$ eV, and 
$U_{\textrm{ring}}=3.27t_{\textrm{ring}}=10.33$ eV, all of which are reasonably close to 10 eV, i.e., the 
value that was used in the previous section. The variation of $U/t$ between the isomers is minor and the consistency
between these results and those of the previous section supports our approach of determining $t$ from DFT calculations.
Using the upper bounds on $U/t$ for the cage and the bowl with the associated minimal values for
the hopping integrals results in $U=8.3$ eV and $U=10.09$ eV, respectively.
It is noteworthy that the value of $(U/t)_{\textrm{cage}}$ is clearly beyond the metal-insulator transition 
point $U/t=4.1$ predicted in 
Ref. \onlinecite{lin07}. This suggests that undoped molecular solids formed of dodecahedral C$_{20}$ are insulators, 
and the isolated molecule is likely not Jahn-Teller active. \cite{lin07} 

In conclusion, the estimation of $U/t$ using the molecular affinity energy in combination with a determination
of the hopping integral $t$ from DFT is a reasonable approach for 
the cage, bowl and ring isomers. We summarize all the isomer parameters in Tab. \ref{parameters}. 
\begin{table}[b]
  \centering
  \begin{tabular}{|cccc|}
    \hline\hline
  & $t$ (eV) & $U$ (eV) & $U/t$ \\
    \hline
 cage & 0.68-1.36   & 8.3-9.67  & 7.1-12.2   \\
    \hline
 bowl & 2.06-2.36   & 10.09-10.15    &   4.30-4.9 \\
    \hline
 ring & 3.16   & 10.33    &   3.27  \\
  \hline\hline
  \end{tabular}
\caption{Hopping $t$ and on-site Coulomb interaction $U$ for C$_{20}$ isomers: cage, bowl and ring.} 
\label{parameters}
\end{table}

\section{Energy gaps and effect of electron doping}\label{sec:doping}
\begin{figure}[t]
\begin{center}
\includegraphics[clip,width=8cm]{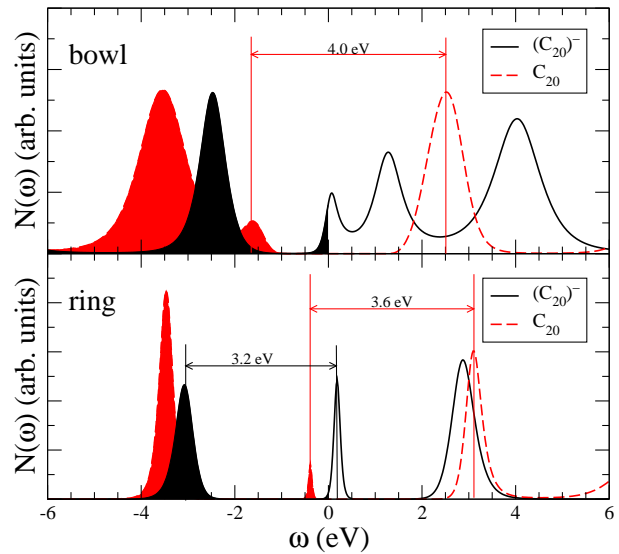}
  \caption{(Color online.) QMC results for the DOS of one-electron doped C$_{20}$ (solid lines) compared with the neutral 
  molecules (dashed lines). For the bowl, $(U/t)_{\rm bowl}=4.30$; 
  for the ring, $(U/t)_{\rm ring}=3.27$. Shaded areas are occupied by electrons. Fermi energies are located at $\omega=0$. 
  Energy units have been converted to eV using $t_{\textrm{bowl}}=2.36$ eV and $t_{\rm ring}=3.16$ eV, respectively. Energy 
  gaps for the neutral molecules and doped ring are shown in the figure.}
  \label{qmcdos_doped}
\end{center}
\end{figure}

We proceed to study the effect of one electron doping on the single particle excitation 
spectra of C$_{20}$ bowl and ring isomers with QMC simulations, using the hopping integrals $t_{\rm bowl}\simeq 2.36$ eV and $t_{\rm ring}=3.16$ eV and 
on-site Coulomb interaction strength $U$ estimated above. A similar study for the fullerene
cage would require time consuming ED calculations which we have not performed. The DOS are shown in 
Fig. \ref{qmcdos_doped}. We see that with one electron doping, spectral 
peaks for both bowl and ring move toward Fermi levels.  Clearly, the one-electron-doped bowl becomes 
metallic. For the one-electron-doped ring, the spectral weight is very close to the Fermi energy. However,
within the precision of the QMC results, we expect the doped ring to remain an insulator
with a gap $\Delta_{\textrm{ring$^-$}}^{\textrm{QMC}}=3.2$ eV.
This is in contrast to 
DFT energy levels, Table \ref{dfthuckel}, which show that with one electron doping both the bowl and the ring isomers
become metallic. 

From these figures we can also estimate 
the energy gaps for the neutral bowl and ring to be around $\Delta_{\textrm{bowl}}^{\textrm{QMC}}=4.0$ eV and 
$\Delta_{\textrm{ring}}^{\textrm{QMC}}=3.6$ eV in reasonable agreement with the rough estimate
given in section~\ref{sec:effect}. We note that these values are much larger than the corresponding DFT 
values $\Delta_{\textrm{bowl}}^{\textrm{DFT}}=2.4$ eV and 
$\Delta_{\textrm{ring}}^{\textrm{DFT}}=1.4$ eV. From the above discussion we expect our previous
estimate of the gap for the fullerene cage of
$\Delta_{\textrm{cage}}=3.1t_{\textrm{cage}}=4.2$ eV at
$(U/t)_{\textrm{cage}}\sim 7$ to be relatively precise. Hopefully, these estimates of the gaps can be compared to
future experimental PES and inverse PES of the neutral C$_{20}$ isomers.

\section{Conclusions}
We proposed an effective Hamiltonian approach to study the electronic correlations in 
C$_{20}$ isomers with QMC simulations and exact diagonalization. The hopping integral, $t$, in an effective one band
Hubbard model Hamiltonian 
are determined by comparing DFT energy levels with the tight-binding Huckel energy 
levels. On-site Coulomb interactions, $U$, are then determined by comparison to the experimental 
affinity energies of the isomers. With these estimated parameters, QMC calculations of the resulting effective
Hubbard model then predicts insulating behavior of the neutral (cage, bowl and ring) and one-electron-doped (ring) isomers and 
metallic behavior of the one-electron-doped bowl isomer. We find qualitative agreement between QMC and DFT calculations for neutral 
(bowl and ring) isomers and one-electron-doped bowl isomer, although QMC gives much larger energy gaps for the neutral isomers. 
For the neutral cage isomer, the QMC prediction (insulating) is qualitatively different from DFT calculations (metallic), since
the cage C$_{20}$ is the most strongly-correlated molecule of the three isomers with a ratio of on-site Coulomb interaction 
and hopping integral of $(U/t)_{\textrm{cage}}\simeq 7.1-12.2$ exceeding the value of $4.1$ for the predicted metal-insulator transition. 
Results presented in the paper await a comparison with 
possible future PES and inverse PES experiments on the gas phase of C$_{20}$ isomers.

\begin{acknowledgments}
FL is supported by the US Department of Energy under award number
DE-FG52-06NA26170. ESS is supported by the Natural Sciences and Engineering
Research Council of Canada and the Canadian Foundation for Innovation.
Computation time were provided by SHARCNET and NCSA supercomputing facilities.
\end{acknowledgments}

\bibliography{c20isomerbib}

\end{document}